\def\be{\begin{equation}}
\def\ee{\end{equation}}
\def\ba{\begin{eqnarray}}
\def\ea{\end{eqnarray}}
\def\l{\label}

\documentclass[12pt]{article}

\usepackage{epsfig}

\begin{document} 
\title{QCD dipole model and $k_T$ factorization}

\author{A.Bialas$^{a,b}$  H.Navelet$^a$ and R. Peschanski$^a$ \\ \\
$^a$Service de Physique Th\'eorique, CEA-Saclay\thanks{Address: SPhT  
CEA-Saclay  
F-91191
Gif-sur-Yvette Cedex, France; e-mail:navelet,  pesch@spht.saclay.cea.fr} 
\\
$^b$M.Smoluchowski Institute of Physics, Jagellonian 
University\thanks{Address: Reymonta 4, 30-059 
Krakow, Poland;
e-mail:bialas@thp1.if.uj.edu.pl}\\Institute of Nuclear Physics, Cracow\\}

\maketitle

\begin{abstract}

It is shown that the colour dipole approach to hard scattering at high
energy is fully compatible with $k_T$ factorization at the leading 
logarithm
approximation (in $-\log x_{Bj}$). The relations between the dipole
amplitudes and unintegrated diagonal and non-diagonal gluon
distributions are given. It is also shown that including  the exact
gluon kinematics in the $k_T$ factorization formula destroys
the
conservation of transverse position vectors and thus is incompatible
with the dipole model  for both elastic and diffractive
amplitudes.

\end{abstract}

\section{ Introduction}

An attractive feature of the colour dipole approach to high-energy
interactions \cite{fra,nz3,nz2} is that it provides a very simple
physical interpretation of the physics at small values of the Bjorken 
variable
$x_{Bj}$. As it can also be justified in the framework of perturbative
QCD in the leading logarithm approximation \cite{mue} ( following the
BFKL approach \cite{bfkl,Li86}), it represents an
interesting possibility for the  description of hard scattering at small
$x_{Bj}$. Indeed, even in its simplest version, the colour dipole model
turned out to be rather successful in describing the total \cite{npr}
and diffractive \cite{bpr} virtual photon - nucleon cross-sections. Its
generalizations are now commonly used for the parametrization of data
from HERA and the Tevatron \cite{gbw}-\cite{dev}.

The basic ingredients of the model  are the dipole-dipole cross-section
and the distribution of dipoles in the projectile and in the
target\footnote{Thus one deals only with colourless objects which gives
a certain theoretical advantage in the extrapolation to the
non-perturbative region.}. It is formulated in transverse
(configuration) space, taking advantage of the fact that at very high
energy the transverse positions of the colliding objects are -
 to a good    approximation - conserved during the collision. This is, in 
fact, the 
origin of the simplicity of the dipole model approach to high-energy
scattering.

It appears important to confront these attractive features of the dipole 
model to the 
present knowledge on perturbative QCD resummations  at leading and 
next-to-leading level in 
logarithms of the energy (i.e. 
$-\log x_{Bj}$). Among the basic known properties of the sum of the 
corresponding  Feynman diagrams, the coupling of external sources (in 
particular a virtual 
photon in deep-inelastic reactions) is  based on the theorem of $k_T$ 
factorization, proven in the 
leading
logarithmic approximation of QCD \cite{cia}. The theorem states that the
``unintegrated'' gluon distribution, i.e. the distribution of energy {\it 
and}
transverse momentum of gluons in the target, factorizes from the rest of
the process. The remaining factor is the so-called ``impact factor''. 
This 
impact 
factor  is an
universal quantity, the same in all processes initiated by the same 
external source, e.g. the photon. 
The ``unintegrated'' gluon distribution characterizes  the target, but 
again the target impact factor 
can also be factorized out, leaving place to an universal interaction 
term, 
given by the BFKL Pomeron 
in the same leading
logarithmic approach  \cite{cia}. At next-to-leading level, the modified 
interaction term is now known \cite{next} but  the impact factors are not 
yet determined.

It should be emphasized that, although both colour dipole model and $k_T$
factorization were (till now) justified only at the leading logarithm 
level,
in practical applications it is necessary to go beyond this approximation 
in order to fix the energy scale of the problem. Before knowing the exact 
next-leading approximation, the current extention beyond the
leading order is  different in the two approaches. In the dipole model 
it is necessary to postulate the relation between $x_{Bj}$ and the energy 
available for the
dipole cascade to develop. In the $k_T$ factorization approach used in 
phenomenology (see 
e.g.\cite{dev}) the relation
between $x_{Bj}$ and gluon longitudinal momentum is fixed by the 
kinematics of the
corresponding Feynman diagram. As neither of these methods can be 
justified 
without 
extensive next-to-leading-order calculations, it remains an open question 
to know either which one describes better the physical reality or how 
both 
are to be modified.

In  the present paper we discuss the relation  between these two
approaches. We start with the   $k_T$
 factorized expression for the total cross-section
(with longitudinal polarization of the photon, but the results extend to 
the transverse one in the 
same way) assuming either full kinematics  or its leading-log 
contribution. 
In the last case we prove 
the exact 
equivalence with the dipole model expression as expected, since they are 
both based 
on BFKL dynamics. In the former one, we show an explicit violation 
of the conservation of transverse positions of the colliding objects  
during the collision.  
 Taking as an explicit example
the diffractive production of two jets, we extend our discussion to the 
off-diagonal gluon 
distributions, where similar differences appear when using the full 
kinematics.

Our main conclusion is that, although equivalent at the leading
logarithm level, the extensions of the dipole model and those based on 
$k_T$ factorization to the next-to-leading
order  lead to  results {\it which are not compatible with each other}. 
This conclusion emphasizes the urgent need for the full next order 
calculation\footnote{Such a calculation is under way \cite{bar}.}
which would settle the question of validity of the two 
most favoured approaches to the  hard collisions at high energy. In 
particular, the fate of  
transverse coordinate conservation in high-energy diffractive collisions
is to be examined. This seems important, as 
some nowadays quite popular models \cite{gbw}-\cite{dev} are based on the 
assumption  that this 
property remains true not only to all orders of the perturbation theory 
but 
even extend to  the 
non-perturbative regime.
 
In the next section we remind briefly the results of the QCD dipole
model for the total (virtual) photon cross-section and show in Section 3
that they are equivalent to the results obtained using $k_T$
factorization in the leading logarithm  approximation. Next, $k_T$
factorization with exact gluon kinematics is discussed. In Section 4 it
is shown to be incompatible with the QCD dipole model. 
 In Section
5 we extend the arguments to inelastic collisions involving the
non-diagonal gluon distributions. Our conclusions are summarized in the
last section.

\section {Total cross-section in the colour dipole approach} 
 
 Let us first remind briefly for future reference the results obtained in
the dipole model for the total cross-section of the virtual photon on an 
arbitrary 
target. The cross-section formula reads (c.f.,
e.g., \cite{b2})
 \be 
\sigma(Y;Q) = 2 N_c \int d^2r dz \ \sigma(r;Y) \  |\Psi(r,z;Q)|^2 
 \l {e1}
 \ee
where the factor $2N_c$ represents summation over spins and colours of
the $q\bar{q}$ states describing the virtual photon, and 
(through the 
optical theorem)
\be
 \sigma(r;Y) = 2 <\vec{r},z|T(t\!=\!0;Y)|\vec{r},z> \equiv  2T(r;Y) 
\l{e2e}
\ee
 is  the dipole-target cross-section and
 $\Psi(r,z;Q)$ is the light-cone photon wave function
(our notation is explicit on Fig.1). It is to be noted that $T$ is a 
function
of only one transverse vector $ \vec{r}$. This is the consequence of the 
diagonal
character of the interaction (following from the conservation of the 
impact 
parameter in
high-energy collisions).

\begin{figure}
\begin{center}
\epsfig{file=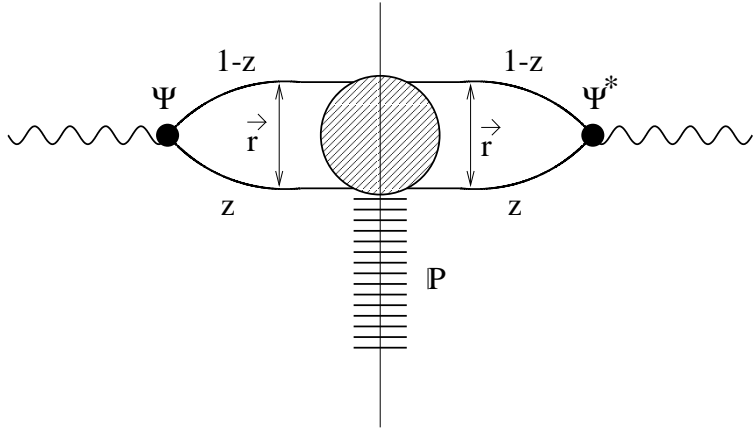,width=10cm}
\end{center}
\caption{Kinematics of the dipole model}
\end{figure}

We shall exclusively discuss  longitudinal
cross-sections\footnote{The  transverse
cross-sections can be treated similarly and the obtained conclusions are
identical.}
 and thus give below only the formula for the wave function of the 
longitudinal photon \cite{bks}:
\be
\Psi_L(r,z;Q) = \frac{ e_q\sqrt{\alpha}}{2\pi}
\ 2z(1-z)QK_0(\hat{Q}r)  \label{11}
\ee
with $\hat{Q}^2= z(1-z)Q^2$.

In case the target is itself a dipole of transverse size $r_0$ (one may 
more 
generally expand the 
target wave-function over a basis of such states), the
cross-section (\ref{e1})
is given by \cite{mp}
\be 
<\vec{r},z|T(t\!=\!0;Y;r_0)|\vec{r},z>=2 \pi
\alpha_s^2r_0^2 \int\frac{d\gamma}{2\pi i} e^{\Delta(\gamma)Y}
\left(\frac{r}{r_0}\right)^{2\gamma}h(\gamma) \l{e2} 
\ee
 where
 \be
\Delta(\gamma) = \frac{\alpha_s N_c}{\pi}
\left[2\psi(1)-\psi\left(1-\gamma\right)
-\psi\left(\gamma\right)\right] \l{e3}
 \ee 
is the well-known BFKL \cite{bfkl} kernel eigenvalue and 
\be
h(\gamma)=\frac1{4\gamma^2(1-\gamma)^2}. \l{e4}
 \ee 
Finally, $Y$ is the
rapidity range available for the dipole cascade to develop.
 $Y$ is not uniquely defined in the context of the dipole
model, a consequence of its origin in the leading logarithm 
approximation\footnote{This is
one of the major  ambiguities in the application of the QCD dipole model 
to
the data \cite{bpr}.}. One expects that $Y$ is of the form
\be
 Y= \log\left(\frac1{x_{Bj}}\right)- Y_0 \l{a6} 
\ee
$Y_0$ can -at least in principle- depend 
on all variables (except $x_{Bj}$) which are relevant in
the process considered. In the case of the total cross-section a
successfull phenomenology \cite{npr} gave $Y_0=const$
but additional dependence on other variables, in particular on $z$ can
also be envisaged \cite{b,bcf}. As emphasized in the Introduction, 
selecting a definite form of
$Y_0$ cannot be justified in the leading logarithm approximation (on 
which 
the dipole model 
is based) and thus represents additional assumption which can only be 
supported by
 phenomenological arguments.

\section {Total cross-section and the $k_T$ factorization.}

We shall now recall the results obtained for the total cross-section
in the approach using the $k_T$ factorization as the starting point,
and show that, when considered in the 
leading-log approximation, they are equivalent to the dipole model 
predictions. As
mentioned in the introduction, we introduce the "unintegrated"
gluon distribution which we shall denote by $g(x_g, k^2)$ where $x_g$ is
the target (nucleon) momentum fraction carried by the gluon and
$\vec{k}$ is the gluon transverse momentum.
\begin{figure}
\begin{center}
\begin{tabular}{cc}
\epsfig{file=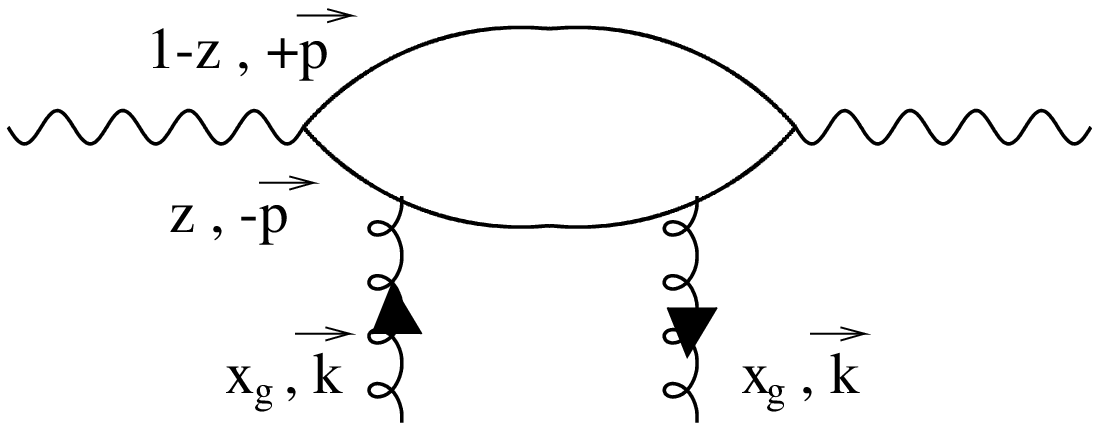,width=6cm} \quad
& \quad \epsfig{file=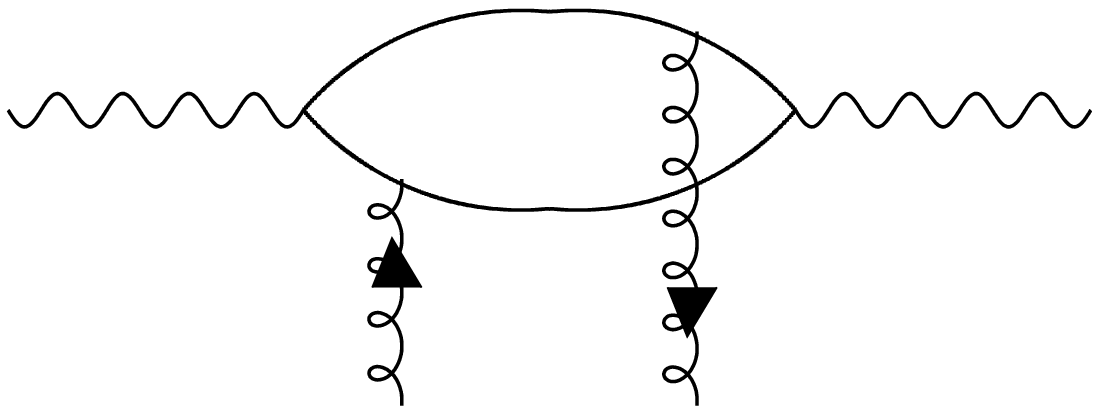,width=6cm}
\end{tabular}
\end{center}
\caption{Kinematics of  two gluon exchange in the $k_T$ factorization 
approach.}
\end{figure}

\bigskip
The cross-section is evaluated from the diagram shown in Fig. 2. The
relevant formulae can be found, e.g., in \cite{kms} where one
reads\footnote{For the derivation, see \cite{for}. To simplify the 
formulae, we neglect the quark 
masses.
Our conclusions do not depend on this simplification (see, e.g., 
\cite{mu}).}

\ba
\sigma_L= \frac{4\pi^2\alpha}{Q^2} F_L=\frac{ 8\alpha \alpha_s e_f^2Q^2 
}{\pi} 
\int_0^1 dz[z(1-z)]^2\nonumber \\ \times \ \int \frac{d^2k}{k^4}  
\int d^2p\left( \frac1{\hat{Q}^2+p^2} - 
\frac1{\hat{Q}^2+(p+k)^2}\right)^2
 g(x_g,k^2) \l{16} 
\ea

Using twice the identity
\be
\frac1{p^2+\hat{Q}^2} = \frac1{2\pi}\int d^2r e^{i\vec{p}\vec{r}} 
K_0(\hat{Q}r)
\l{x1}
\ee
results in the following expression:
\ba
\sigma_L= \frac{ 8\alpha \alpha_s e_f^2Q^2 }{\pi}\frac1{(2\pi)^2}\int_0^1 
dz[z(1-z)]^2
 \int d^2r  K_0(\hat{Q}r)\int d^2r'  K_0(\hat{Q}r') \nonumber \\
\times \ \int d^2p \ e^{i\vec{p}(\vec{r}-\vec{r'})} D(\vec{r},\vec{r'})   
\l{x2}
\ea
where
\ba
D(\vec{r},\vec{r'})=
 \int \frac{d^2k}{k^4}g(x_g,k^2) \left(1-e^{-i\vec{k}\vec{r}}\right)
 \left(1-e^{i\vec{k}\vec{r'}}\right).   \l{x3}
\ea

The formulae (\ref{x2}) and  (\ref {x3}) provide  the key point of our
discussion. The conservation of transverse position throughout the
interaction would mean that
\be
 \int d^2p
\ 
e^{i\vec{p}(\vec{r}-\vec{r'})}D(\vec{r},\vec{r'}) \propto
\delta^2(\vec{r}-\vec{r'})\ .  \l{x3g}
\ee
As this may happen only if $D(\vec{r},\vec{r'})$ does not depend on
$\vec{p}$, 
it becomes clear from (\ref {x3}) that the question of
the gluon kinematics, i.e. the determination of $x_g$ as a function of
the other variables, is crucial.  Indeed, if the gluon kinematics is 
 taken into account, $x_g$ can be expressed by other variables (see, 
e.g.\cite{kms}):
\be 
x_g= x_{Bj}\cdot \ \frac{\hat{Q}^2+\hat{k}^2+ (p')^2}{\hat{Q}^2} \ ;\;\; 
\vec{p'}=\vec{p}
- (1\!-\!z) \vec{k}\ .
\l{y1} 
\ee 
Consequently, as seen from (\ref{x3}), $D(\vec{r},\vec{r'})$ does depend
on $\vec{p}$. Postponing the detailed discussion of this case to the
next section, we shall now consider the formula (\ref{x3}) in the
leading logarithm approximation (for which the $k_T$ factorization can
be exactly proven). In this approximation the relation between $x_{Bj}$
and $x_g$ is not well defined because in the limit $x_{Bj} \rightarrow
0$, the difference between $\log x_{Bj}$ and $\log x_g$ is finite and
thus any contribution depending on this difference (or -equivalently- on
the ratio $x_{Bj}/x_g$) must be of higher order. Taking this ambiguity
into account we can treat $x_g$ in (\ref{16}) as a constant, independent
of other variables entering the process\footnote{This property was
actually used in the proof of the $k_T$ factorization \cite{cia}.}.
Under this condition 
 (\ref{16}) can be transformed into the
form given by the dipole model formula discussed in the previous
section \cite{nz2,baro}.

Indeed, since $D(\vec{r},\vec{r'})$ does not depend on $\vec{p}$, the 
integral
over $d^2p$ gives $(2\pi)^2\delta^2(\vec{r}-\vec{r'})$  and thus we 
obtain
\ba
\sigma_L=  \frac {16 \pi\alpha \alpha_s e_f^2Q^2}{\pi}\int_0^1 
dz[z(1-z)]^2
 \int d^2r  [K_0(\hat{Q}r)]^2  \nonumber \\
\times \ 
\int \frac{d^2k}{k^4}g(x_g,k^2) \left(1-e^{-i\vec{k}\vec{r}}\right)
\ .
   \l{xx2}
\ea
Using (\ref{11})  this can be written in the form of (\ref{e1}) with the 
 function ${T}(r,Y)$  now given by \cite{baro}
\be
{T}(r,Y)=\frac{2\alpha_s \pi}{N_c}\int \frac{d^2k}{k^4}g(x_g,k^2)
\left (1-e^{i\vec{k}\vec{r}}\right) \ .  \label{19}
\ee
This formula  expresses the dipole-target forward elastic amplitude in
terms of the unintegrated gluon distribution in the target. It
 can be inverted. To this end we calculate the moments of $T$.
Using (\ref{19}) and the identity
\be
\int_0^{\infty} \frac{du}{u^{1+\gamma}} [1-J_0(u)] = \frac
{\Gamma(\gamma/2)\Gamma(1-\gamma/2)}{2^{1+\gamma}[\Gamma(1+\gamma/2)]^2}
\l{22}
\ee
one obtains
\be
\tilde{T}(\gamma,Y) \equiv \frac1{2\pi}\int \! d^2r \ r^{-2-2\gamma} \ 
T(r,Y)
= \frac{4\pi^2\alpha_s}{N_c}\left\{\frac
{\Gamma(\gamma)\Gamma(1-\gamma)}{2^{1+2\gamma}[\Gamma(1+\gamma)]^2}\right
\}
\tilde{g}(x_g,\gamma)  \l{22a}
\ee
where
\be
\tilde{g}(x_g,\gamma)\equiv \frac1{2\pi}\int \frac{d^2k}{k^4}k^{2\gamma} 
g(x_g,k^2)
=\int \frac{dk}{k}k^{-2+2\gamma} g(x,k^2)
 \l{22b}
\ee 
and the factor between brackets $\left\{...\right\}$
can be interpreted as the gluon-dipole vertex function \cite{mu}.

Using the inverse Mellin transform and (\ref{22a}), we thus obtain
\ba
g(x_g,k^2)\equiv 2\int \frac{d\gamma}{2\pi i}
 \ k ^{2-2\gamma} \ \tilde{g}(x_g,\gamma)
\nonumber \\
= \frac{N_c}{2\pi^2\alpha_s }\int \frac{d\gamma}{2\pi i} \ k ^{2-2\gamma} 
\frac {2^{1+2\gamma}[\Gamma(1+\gamma)]^2}
{\Gamma(\gamma)\Gamma(1-\gamma)}\ \tilde{T}(\gamma,Y)   \l{22d}
\ea
where $\tilde{T}(\gamma,Y)$ is given by (\ref{22a}).

This is a general formula valid for any target. If the target is itself a
 QCD dipole,
 we can use the formula (\ref{e2}) for $T$ and obtain
\ba
g(x_g,k^2)=\frac{4N_c \alpha_s}{\pi} \int \frac{d\gamma}{2\pi i}
\left(\frac{kr_0}2\right)^{2-2\gamma} 
 h(\gamma)\frac
{[\Gamma(1+\gamma)]^2}{\Gamma(\gamma)\Gamma(1-\gamma)}
\ e^{\Delta(\gamma) Y}.
 \l{26}
\ea

\section{Exact gluon kinematics }

Although the validity of the $k_T$ factorization was established only at
the leading order in $-\log x_{Bj}$, it is natural to treat the diagrams
of Fig. 2 as giving the correct structure of the process (to all
orders). This  is usually the way $k_T$ factorization is
applied for the description of high-energy collisions 
\cite{gbw}-\cite{dev},\cite{kms}.
It should be realized, however, that such an approach takes into
account only part of the higher order corrections and thus cannot be
justified from the first principles without estimating contributions
from other diagrams which appear at higher orders.

The important consequence of this procedure is, in particular, that the
kinematics of the diagram is to be fully taken into account and thus the 
value of
$x_g$ in Eq.(\ref{16}) becomes now well-defined and given by
(\ref{y1}). As we have already indicated in the previous
section, it turns out that these relations break the connection between
the formula (\ref{16}) of $k_T$ factorization and (\ref{e1}) of the
colour dipole model, established in Section 3 at the leading logarithm
level. Indeed, in this case we have 
\ba
D(\vec{r},\vec{r'})= \int
\frac{d^2k}{k^4}g\left(x_{Bj} \frac{\hat{Q}^2+\hat{k}^2+
(p')^2}{\hat{Q}^2} ,k^2\right) \left(1-e^{-i\vec{k}\vec{r}}\right)
\left(1-e^{i\vec{k}\vec{r'}}\right). \l{y3} 
\ea
Thus $D$ does depend on $\vec{p}$ and, consequently, the integral over 
$d^2p$ in (\ref{x3g}) does not give $\delta^2(\vec{r}-\vec{r'})$ (except
in the particular case when $g(x_g,k^2)$ does not depend on $x_g$ at
all, corresponding to the energy-independent cross-section).
To obtain a better insight, we express the gluon distribution in the form 
of
a Mellin transform:
\be
g(x_g,k^2)=\int\frac{dN}{2i\pi} (x_g)^{-N} g_N(k^2)    \l{y4}
\ee
and find (see \cite{gra})
\ba
\int d^2p\  e^{i\vec{p}'(\vec{r}-\vec{r'})}D(\vec{r},\vec{r},) \propto
\int \frac{dN}{2i\pi} g_N(k^2) \frac{(\hat{Q}^2)^N}{(2x_{Bj})^N 
\Gamma(N)}
\nonumber \\ K_{1-N}
\left(|\vec{r}-\vec{r'}|\sqrt{\hat{Q}^2+\hat{k}^2} \right)
\left|\frac{\sqrt{\hat{Q}^2+\hat{k}^2}}{\vec{r}-\vec{r'}}\right|^{1-N}
\ 
.
\l{y4y}
\ea

One thus  explicitly sees 
that the process is no longer diagional in
transverse size $\vec{r}$ of the dipole. Since the diagonal character of
the dipole interaction in the transverse space is the fundamental
ingredient of the colour dipole approach, we must conclude that {\it the
 extension of the theorem of $k_T$ factorization \cite{dev,kms}
to non-leading order by including the exact
gluon kinematics is not compatible with the colour dipole model}.
This conclusion seems important, as it  emphasizes the need for a better
understanding of the standard phenomenology of hard high-energy
interactions.

\section{ Diffractive processes: off-diagonal gluon distributions}  

To complete the argument, we shall  discuss also inelastic diffraction.
The simplest example for which results were presented in both models 
is the two jet production off any target.

In the dipole model the cross-section for this process can be readily
obtained taking as a starting point the formula for quasi-elastic
diffraction given in \cite{bp}. The result is
\be
\frac{d\sigma}{dz d^2p dt} = \frac{N_c}{2\pi} \ |G(\hat{p},z,t;Y^*)|^2 
\l{a1}
\ee
where
\be
G(p,z,t;Y^*)=\frac1{2\pi} \int d^2r \exp(i\vec{p}\ . \ \vec{r})
<\vec{r},z|T(t;Y^*)|\vec{r},z> \Psi(\vec{r},z;Q)\ .  \l{a2}
\ee
Here $\vec{p}$ is the transverse momentum of the jet and 
 $\Psi(\vec{r},z;Q)$ is the photon wave
function given by (\ref{11})\footnote{ As before,  we shall discuss 
explicitly only 
longitudinal
photons. The case of transverse photons can be treated on the same 
lines.}.

In the case when the target is itself a dipole of size $r_0$,
the  amplitude
$<\vec{r},z|T(t;Y^*;r_0)|\vec{r},z>$ was first discussed in \cite{lip}
and explicitly calculated in \cite{bnp}. Here we shall restrict ourselves 
to 
the forward scattering $t= 0$ \cite{bc}. In this case the formula for 
$T$ is much simpler and given by (\ref{e2}), up to the determination of 
$Y^*$.

As already explained, $Y^*$ is the rapidity range available for the 
dipole
cascade to develop. It should be emphasized again that it remains 
undetermined
and, in particular, need not  be the same as in (\ref{e2}).
The general formula (\ref{a6}) remains valid but in the case of the 
diffractive production a
successful phenomenology \cite{bpr}
 required (see formula (\ref{a6}))
 $Y_0=\log \beta \equiv \log \frac {Q^2}{Q^2+M^2}$, i.e.
\be
Y^*= \log\left(\frac{1}{x_{P}}\right);\;\;\;\; x_{P}=x_{Bj}/\beta. 
\l{aa7}
\ee

We shall now show that, at the leading logarithm approximation, the
formula for the diffractive two-jet production obtained with the help
the $k_T$ factorization can also be expressed in the form
(\ref{a1},\ref{a2}) derived from the dipole model. We also derive an 
explicit 
relation between the dipole elastic amplitude and the off-diagonal 
gluon distribution.

 We start from  the formula of \cite{gkm}, giving the forward ($t\!=\!0$) 
differential
cross-section for diffractive production of two jets 
(see also  \cite{nz,blw})
\ba
\frac{d\sigma_L}{dtdzd^2p }\! = \!\frac{\alpha \alpha_s^2}{2\pi N_c}
e_q^2 \int\frac{d^2k d^2k'}{k^4k'^4} f(x_{Bj},x'\!,x''\!,k^2) 
f(x_{Bj},x'\!,x''\!,k'^2)
\Phi_L(z,k,k'\!,p) \label{1}
\ea
where the function $f(x_{Bj},x',x'',k^2)$ is the linear combination of 
the 
off-diagonal
\cite{rad,ji}
gluon distributions inside a proton
\be
f(x_{Bj},x',x'',k^2) = \frac12 
\left\{g(x'+x_P,x',k^2)+g(x''+x_P,x'',k^2)\right\}  
\label{2}
\ee
where $x'$ and $x''$ are  the gluon momenta as explained in \cite{gkm} 
 and $x_P$ given by (\ref{aa7}). 

Since the expression for
 $\Phi_L$ (given in \cite{gkm}) can be written in the factorized form
\ba
\Phi_L =4Q^2z^2(1\!-\!z)^2 \left[\frac1{p^2\!+\!\hat{Q}^2} 
-\frac1{(p\!+\!k)^2\!+\!\hat{Q}^2}\right]
\left[\frac1{p^2\!+\!\hat{Q}^2} 
-\frac1{(p\!+\!k')^2\!+\!\hat{Q}^2}\right]  
\label{5}
\ea
where $\hat{Q}^2=z(1\!-\!z)Q^2$, one sees immediately that (\ref{1}) can 
be 
cast in the form (\ref{a1}) with
\ba
G\!=\!\frac{\alpha_s\sqrt{\alpha_{em}e_q^2}}{N_c} 
 2Qz(1\!\!-\!\!z)\int \frac{d^2k}{k^4}f(x_{Bj},x'\!,x''\!,k^2)
\left[\frac1{p^2+\!\hat{Q}^2}\! 
-\!\frac1{(p\!+\!k)^2\!+\!\hat{Q}^2}\right]\ 
.\nonumber \\  
\label{6}
\ea
It remains to be shown that $G$ given by (\ref{6}) can be written in the 
form
(\ref{a2}).

To see this, we follow closely the argument of Section 3. Using the
identity (\ref{x1}) we write the first term of (\ref{6}) in the form of
an integral over $d^2r$:
\be
\int \frac{d^2k}{k^4} f(x_{Bj},x'\!,x''\!,k^2) \frac1{p^2\!+\!\hat{Q}^2} 
=
\frac1{2\pi}\int d^2r e^{i\vec{p}\ \vec{r}} \int \frac{d^2k}{k^4}
f(x_{Bj},x'\!,x''\!,k^2) K_0(\hat{Q} r) \ . \l{8}
\ee
The same operation can be applied to the second term 
and thus we obtain
\be
G=\frac1{2\pi} \int d^2r \ e^{ipr}\ T(r,z,x,x_P)\ \Psi_L(r,z;Q)
\label{10}
\ee
where $\Psi_L$ is given by (\ref{11})
and 
\ba
T(r,z,x,x_P)=\frac {2\pi}{N_c}\alpha_s 
\int \frac{d^2k}{k^4}\left(1- e^{i\vec{k}\vec{r}}\right)
f(x_{Bj},x',x'',k^2) \ .     \label{12}
\ea
Thus we recover a similar formula as in the case of total cross-section
(see Section 2, Eq.(\ref{19})),
except that now the unintegrated gluon distribution $g(x,k^2)$ is
replaced by the unintegrated  {\it off-diagonal}  gluon distribution
$f(x_{Bj},x',x'',k^2)$,
 as appropriate for inelastic processes \cite{rad,ji}.

One sees that the Eq.(\ref{12}) is valid even if exact gluon kinematics
 is included (at no place in the
derivation we needed the assumption that $x'$ and $x''$ are constants,
independent of $\vec{p}$). Its physical interpretation, however, depends
crucially on this problem. Indeed, if $x'$ and $x''$ do not depend on
$\vec{p}$, also $T$ given by (\ref{12}) is $\vec{p}$-independent and {\it 
only
in this case it can be interpreted as the amplitude for scattering of the
dipole of a given transverse size $\vec{r}$}. If, on the other hand, the
exact gluon kinematics is taken into account (which includes partly
contribution of higher orders) one has \cite{gkm}
\be
x'= x_P\frac {k^2 +2\vec{p}\cdot \vec{k}}{z(M^2+Q^2)};\;\;\;\;x''
= x_P\frac {k^2 +2\vec{p}\cdot \vec{k}}{(1-z)(M^2+Q^2)}   \label{3}
\ee
and thus $T$ is a function of both $\vec{r}$ and $\vec{p}$. This of
course cannot be the case  if it is to be interpreted as the
scattering amplitude.

Thus we conclude that, similarly as in the case of total cross-section,
the description of diffractive scattering in the dipole model is only
compatible with the $k_T$ factorization if one restricts to the leading
logarithm level.

If one  stays at the leading logarithm approximation,
one can of course repeat the argument of Section 3 and thus conclude
 that the  relations between
the dipole amplitude and the gluon distribution  of
 Section 3 should be  valid  with the replacement
$g\leftrightarrow f$ and $Y\leftrightarrow Y^*$. In particular 
\be
f(x_{Bj},x',x'',k^2)= 
\frac{N_c}{2\pi^2\alpha_s }\int \frac{d\gamma}{2\pi i} \ k ^{2-2\gamma} 
\ \frac {2^{1+2\gamma}[\Gamma(1+\gamma)]^2}
{\Gamma(\gamma)\Gamma(1-\gamma)}\ \tilde{T}(\gamma,Y^*) \ .  \l{14a}
\ee
Using (\ref{e2}), one can also write down the explicit 
prediction of the dipole model for the gluon distribution in the 
dipole target:
\ba
f(x_{Bj},x',x'',k^2) = \frac{4N_c}{\pi} \alpha_s
\int\frac{d\gamma}{2\pi i}\left(\frac{kr_0}2\right)^{2-2\gamma}
e^{\Delta(\gamma)Y^*}
 h(\gamma)\frac
{[\Gamma(1\!+\!\gamma)]^2}{\Gamma(\gamma)\Gamma(1\!-\!\gamma)}.  \l{14}
\ea

\section{Discussion and conclusions}

In conclusion, we have shown that {\it in the leading logarithm 
approximation} 
the $k_T$ factorization and the colour dipole model give 
equivalent description of hard
processes at high energy.

On the other hand, when one steps beyond the leading logarithmic 
approximation
by supplementing the $k_T$ factorization algorithm with the 
exact kinematics of the corresponding Feynman diagrams (as is the case in
practical applications \cite{dev}), the result is {\it incompatible} with 
the colour
dipole model: we have shown  that such a procedure leads to a violation 
of 
the 
conservation of transverse positions and sizes of the colliding objects 
(which
is fundamental for the dipole model interpretation of high-energy 
collisions).

This breaking of diagonality of the interaction in impact parameter 
should 
not
be surprising: indeed, since the transverse momentum and impact parameter 
are
conjugate variables, conserving one of them in the interaction does not 
allow to
conserve the other one. In fact, this effect is analogous to that 
discussed
already in \cite{bp}.

Several comments can be made.

(i) As both the colour dipole model and $k_T$ factorization (with exact
kinematics) are currently used for the description of data \cite{dev}, 
our
result indicates that such analyses must be taken with some care. 

(ii) It also emphasizes  the need for a  complete next-to-leading order
calculation which would elucidate the problem of compatibility of the two
approaches and also provide the necessary information on the correct 
energy
scale.

(iii) In absence of such a complete calculation, one may only speculate 
about
the origin of the difficulty. One possibility is that, by including other 
missing higher order contributions, one recovers the compatibility 
between 
the
two pictures. This would mean that the theorem of $k_T$ factorization 
should be
replaced by a sort of "impact parameter factorization".  Another possible
scenario is that $k_T$ factorization  at higher orders will not 
correspond 
to impact parameter
conservation and thus the
dipole model must be abandoned or reformulated.

(iv) Speculating about the possibilities of modification of the colour 
dipole
model, one may think necessary to add (at the next-to-leading order) 
the contributions $ 1\; dipole 
\rightarrow 2\; dipoles$  with the two new dipoles having energies of the 
same
order of magnitude. Such a contribution  would certainly require a 
re-interpretation
of the conservation of transverse positions in high-energy scattering and 
thus
make more plausible the compatibility with $k_T$ factorization.

\vspace{0.3cm}
{\bf Acknowledgements}
\vspace{0.3cm}

A.B. thanks the Service de Physique Theorique of Saclay for support and
kind hospitality. This investigation was supported in part by the KBN
Grant No 2 P03B 086 14 and by the Subsidium of Fundation for Polish
Science 1/99.

\end{document}